\documentclass[12pt]{iopart}
\newcommand*{\ketbra}[1]{\ket{#1}\bra{#1}}
\newcommand{\beq}{\begin{equation}}
\newcommand{\eeq}{\end{equation}}
\newcommand{\beql}[1]{\begin{equation}\label{eq:#1}}
\newcommand{\beqa}{\begin{eqnarray}}
\newcommand{\eeqa}{\end{eqnarray}}
\newcommand{\beqas}{\begin{eqnarray*}}
\newcommand{\eeqas}{\end{eqnarray*}} 
\newcommand{\Eq}[1]{Eq.~(\ref{eq:#1})}              
\newcommand*{\vl}{\vec{l}}
\newcommand*{\vu}{\vec{u}}
\newcommand*{\vv}{\vec{v}}
\newcommand*{\vw}{\vec{w}}
\newcommand*{\vs}{\vec{\sigma}}
\newcommand{\us}{U_{\mathbf{S}}}
\newcommand{\ls}{L_{\mathbf{S}}}
\newcommand{\xs}{X_{\mathbf{S}}}
\newcommand{\ys}{Y_{\mathbf{S}}}
\newcommand{\zs}{Z_{\mathbf{S}}}
\newcommand{\is}{I_{\mathbf{S}}}
\newcommand{\ia}{I_{\mathbf{A}}}
\newcommand{\ib}{I_{\mathbf{B}}}
\newcommand{\li}{l_{{\mathbf{S}},i}}
\newcommand{\lj}{l_{{\mathbf{S}},j}}
\newcommand{\lk}{l_{{\mathbf{S}},k}}
\newcommand{\lon}{l_{{\mathbf{S}},1}}
\newcommand{\ltw}{l_{{\mathbf{S}},2}}
\newcommand{\lth}{l_{{\mathbf{S}},3}}
\newcommand{\bA}{{\bf A}}
\newcommand{\bB}{{\bf B}}

\newcommand{\bS}{{\bf S}}

\newcommand{\cE}{{\cal E}}
\newcommand{\cH}{{\cal H}}
\newcommand{\al}{\alpha}
 
\newcommand{\ps}{\psi}

\newcommand{\nol}{\nonumber\\}

\newcommand{\ph}{\phi}

\newcommand{\Ps}{\Psi}

\newcommand{\si}{\sigma} 
\newcommand{\da}{\dagger}
\newcommand{\eq}[1]{(\ref{eq:#1})}
\newcommand{\bra}[1]{\langle#1|}
\newcommand{\ket}[1]{|#1\rangle}

\newcommand{\bracket}[1]{\langle#1\rangle}

\newcommand{\cut}[1]{}

\usepackage{iopams}  
\usepackage{graphicx}
\usepackage{graphicx,color}

\begin{document}
\def\av#1{\langle#1\rangle}
\title[Gate fidelity of arbitrary single-qubit gates constrained by conservation laws]
{Gate fidelity of arbitrary single-qubit gates constrained by conservation laws}

\author{Tokishiro Karasawa${}^{1}$, Julio Gea-Banacloche${}^{2}$, \\
and Masanao Ozawa${}^{3}$}

\address{${}^{1}$National Institute of Informatics, 
Chiyoda-ku, Tokyo,  101-8430, Japan \\
${}^{2}$Department of Physics, University of Arkansas,
Fayetteville, Arkansas, 72701, USA\\
${}^{3}$Graduate School of Information Science, 
Nagoya University, Chikusa-ku, Nagoya, 464-8601, Japan}
\ead{jidai@nii.ac.jp, jgeabana@uark.edu, and ozawa@is.nagoya-u.ac.jp}

\begin{abstract}

Recent investigations show that conservation laws limit
the accuracy of gate operations in quantum computing. 
The inevitable error under the angular momentum
conservation law has been evaluated so far for the
CNOT, Hadamard, and NOT gates for spin 1/2 qubits,
while the SWAP gate has no constraint. 
Here, we extend the above results to general single-qubit gates. 
We obtain an upper bound of the gate fidelity of arbitrary single-qubit gates implemented under arbitrary conservation laws, determined by the geometry of the conservation law and the gate operation on the Bloch sphere as well as the size of the ancilla.

\end{abstract}


\maketitle

\section{Introduction}

One of the most demanding factors in realizing scalable quantum 
computers is the accuracy requirement for implementing the elementary 
quantum gates set by the threshold theorem for successful concatenated 
error-correction \cite{NC00}.
Two types of errors under current consideration are the environment-induced
decoherence, caused by the interaction with the environment, 
and the controller-induced decoherence, caused by the interaction
with the controller of the gate operation (the latter considered separately from, and in addition to, any errors arising from classical control imperfections).   
The environment-induced decoherence may be
overcome, in principle, by developing qubits with long decoherence
time.  In the early treatments \cite{NC00}, the controller-induced decoherence
was not distinguished from the environment-induced one.  However, 
they have different constraints, since insensitivity and controllability 
are two contradictory demands.

In the conventional description, the controller is described as a classical
system and then causes no decoherence.
However, the quantum nature of electromagnetic control fields has been 
studied by one of the present authors and others 
\cite{BW99,EK02,Ban02,Ban02b,BK03,BK05,SD04,XYS06}, 
and shown to be a potentially substantial source of decoherence; see also 
Refs.~\cite{Ita03,EK03,Ban03} for a debate on the validity of the
model under consideration.

In contrast to the above model-dependent approach,
one of the present authors \cite{02CQC,02CLU,03UPQ,03QLM} 
independently found physical constraints on gate operations generally 
imposed by conservation laws, 
by quantitatively generalizing the so-called Wigner-Araki-Yanase theorem 
\cite{Wig52,AY60,Yan61}. 
The inevitable error probability under the angular momentum conservation 
law has been shown to be inversely proportional to the variance of 
the controller's conserved quantity for the CNOT gate \cite{02CQC} and 
the Hadamard gate \cite{03UPQ}, while the SWAP gate obeys no constraint \cite{02CQC}.  
For the NOT gate a similar lower bound has been obtained more recently 
by a different method \cite{07CLI}. 
Subsequently, the above two approaches have been compared and 
merged \cite{05CQL,06MEP}.
It was shown there that the limit derived from the angular momentum conservation 
law is equivalent to the one yielded by the phase fluctuations of the control field 
when a qubit-field interaction Hamiltonian of the Jaynes-Cummings type is assumed.

Here, we extend the above results to arbitrary single-qubit gates
with arbitrary conservation laws from a geometrical point of
view.  
We show that a lower bound of the gate infidelity, one
minus the squared gate fidelity, is given by
\begin{equation}
\frac{\sin^2\frac{\theta}{2} \sin^2\Ps (1-\sin^2\frac{\theta}{2} \sin^2\Ps)}{4(1+\si(L/c)^2)},
\label{eq1}
\end{equation}
where $\Psi$ is the relative angle between the axes of
rotations on the Bloch sphere generated by the conserved
quantity and the gate operation; 
 $\theta$ is the rotation angle of the gate in the Bloch sphere, or the difference of the arguments of two eigenvalues of the gate, as defined later
(in particular, $\theta=\pi$ for self-adjoint gates); $L$ is the conserved quantity in the controller; and $c$ is the maximum standard deviation of the conserved quantity in the qubit, while the standard deviation $\si(L/c)$ of $L/c$ is generally understood as the size of the controller.  
If the control interaction has full rotation symmetry, 
we eventually conclude that the gate infidelity of any implementation of an arbitrary gate on a spin 1/2 qubit controlled by a spin N/2 ancilla system is bounded by $(\sin^2\theta)(4+4N^{2})^{-1}$ for $0\leq \theta \leq \pi/2$ and $(4+4N^{2})^{-1}$ for $\pi/2 \leq \theta \leq \pi$ under the angular momentum conservation law. On the other hand, if the control interaction does not have the full rotational symmetry,  the bound (\ref{eq1}) holds.  Because this bound vanishes 
in the important case $\theta = \pi$ and $\Ps=\pi/2$ (corresponding to a NOT, or bit-flip, gate, when the conserved quantity is proportional to the $z$ component of the angular momentum),  we have also developed here an approach that yields an alternative bound that is more suitable than (\ref{eq1}) under some circumstances; this alternative bound scales as  $(1+\sqrt{1+\si(L/c)^2})^{-2}$ with the size of the controller.

Although in many practical cases what we have called the controller will, by its very nature, be too large for our limits to represent a significant constraint, we note that, as pointed out above, laser pulses interacting with atomic qubits may be an exception to this \cite{Ban02b}, especially in view of the recent result that minimum energy pulses cannot be, in general, safely shared or reused \cite{06MEP}.  Other systems for which our results may be quite relevant are the so-called ``programmable quantum processors'' \cite{nielsenchuang,hillery}, which consist of a set of data qubits undergoing joint, closed evolution with a set of ``program'' qubits.  If the total system evolution obeys a conservation law, our results clearly imply that the set of program qubits needs to be sufficiently large, in order to carry out the desired operations to a sufficiently large degree of accuracy.
  
Our paper is organized as follows. Section 2 gives the formulation of the problem and examples 
to which our formulation applies. Section 3 introduces
the deviation operator and gives a lower bound of the infidelity in
terms of the variance of the deviation operator.
In Section 4, we consider the commutator of the deviation
operator and the conserved quantity to obtain a lower bound on
the variance of the deviation operator using Robertson's inequality.  
In Section 5,  the lower bound 
is represented in terms of the relative angle mentioned above
and the size of the controller.
Section 6 considers the case of the full rotational symmetry, while
Section 7 gives an alternative lower bound which is useful for the case
where the above bound vanishes.

\section{Basic formulation}
\label{se:BF}

Let $\bS$ be a qubit with computational basis $\{ \ket{0}, \ket{1} \}$.
An arbitrary single-qubit gate is an arbitrary unitary operator 
$U_{\bS}$ on the Hilbert space $\cH_\bS$ of $\bS$.
The class of arbitrary single-qubit gates includes the Pauli operators defined by $X_{\bS}=\left| 0 \right\rangle\left\langle 1 \right| + \left| 1 
\right\rangle\left\langle 0 \right|$, $
Y_{\bS}=- i \left| 0 \right\rangle\left\langle 1 \right| + i \left| 1 
\right\rangle\left\langle 0 \right|$, and 
$Z_{\bS}=\left| 0 \right\rangle\left\langle 0 \right| - \left| 1 
\right\rangle\left\langle 1 \right|$, as well as the Hadamard gate defined by 
$H_{\bS}= (\sqrt{2})^{-1}\left( 
\left| 0 \right\rangle\left\langle 0 \right| + \left| 0 \right\rangle\left\langle 
1 \right|
+\left| 1 \right\rangle\left\langle 0 \right| - \left| 1 \right\rangle\left\langle 
1 \right|
\right)$.
We introduce a vector $\vs$ defined by $\vs=(\xs,\ys,\zs)$ for convenience in later analysis. The vector $\vs$ provides the general description of an arbitrary Hermitian operator of $\bS$
\beqa
A= \phi \is + \frac{\theta}{2} \vec{u} \cdot \vs,
\label{eq2}
\eeqa
where $\vec{u}$ is a unit vector defined as $\vec{u}=(u_{x},u_{y},u_{z})$ with $\|\vec{u}\|^2={u_{x}}^2+{u_{y}}^2+{u_{z}}^2=1$, and $\phi$ and $\theta$ are real numbers, and $\is$ stands for the identity operator of $\bS$.  Here we have introduced an inner product between an arbitrary (c-number) vector $\vec u$ and the Pauli operators as 
\beqa
\vec{u} \cdot \vs = u_{x}\xs + u_{y}\ys + u_{z}\zs.
\eeqa
As an arbitrary unitary operator can be written as $e^{iA}$, with $A$ Hermitian, we can write an arbitrary single-qubit gate $\us$ as
\beqa
\us =e^{iA} = e^{i\phi}\left(\cos\frac{\theta}{2} \is + i \sin\frac{\theta}{2} \vu\cdot\vs \right).
\label{eq4}
\eeqa
Since two eigenvalues of $U_S$ are $\exp[i(\ph\pm\theta/2)]$, parameters $\phi$ and $\theta$ are uniquely determined with $0\le \phi < 2\pi$ and $0\le \theta \le\pi$.  The angle $\theta$ corresponds to the rotation angle in Bloch sphere induced by the gate $U_S$; see p. 175 of Ref.~\cite{NC00}.

Suppose we want to implement the arbitrary single-qubit gate $\us$ by letting the system $\bS$ interact for a finite time interval with an ancilla $\bA$, which is described as a quantum system with a Hilbert space $\cH_{\bA}$. The total system to be considered is the composite system $\bS+\bA$, and we assume that its initial state is prepared as a product state before the interaction. We want the desired gate $\us$ to be implemented by the time evolution of the composite system. Thus, every possible implementation is characterized by a pair $\al=(\rho_{\bA},U)$ consisting of a density  operator $\rho_{\bA}$ on $\cH_{\bA}$ describing the initial state of $\bA$ and a unitary operator $U$ on $\cH_{\bS}\otimes\cH_{\bA}$ describing the time evolution of $\bS+\bA$
during the interaction \cite{02CQC}. 
An implementation (characterized by) $\al=(\rho_{\bA},U)$ defines a trace-preserving quantum
operation $\cE_{\al}$ by
\beqa
\cE_{\al}(\rho_{\bS})=\Tr_{\bA}[U(\rho_{\bS}\otimes \rho_{\bA})U^{\dagger}]
\eeqa
for any density operator $\rho_{\bS}$ of the system $\bS$, 
where $\Tr_{\bA}$ stands for the partial trace over $\cH_{\bA}$.
For simplicity, in what follows we will assume that $\rho_{\bA}$ is a pure state $\ketbra{A}$, and in this case, the implementation will be described as $\alpha = (\ket{A}, U)$ (the case in which $\rho_{\bA}$ is an arbitrary mixed state is considered in the Appendix). 

How successful the implementation $\al=(\ket{A},U)$
has been is measured by the gate fidelity 
of $\cE_{\al}$ relative to $U_{\bS}$ (Ref.~\cite{NC00}, p.~418) defined by
\beqa\label{eq:fidelity}
F(\cE_{\al},U_{\bS})
=\inf_{\ket{\ps}}F(\ket{\ps}),
\eeqa
where $\ket{\ps}$ varies over all state vectors of $\bS$, and 
$F(\ket{\ps})$
is the fidelity of the two states 
$U_{\bS}\ket{\ps}$ and $\cE_{\al}(\ket{\ps}\bra{\ps})$, given by
\beqa\label{fidelity}
F(\ket{\ps})=
\bracket{\ps|U_{\bS}^{\da}\cE_{\al}(\ket{\ps}\bra{\ps})U_{\bS}|\ps}^{1/2}.
\eeqa

The implementation is perfect, i.e., $\cE_{\al}(\rho)=U_{\bS}\rho U^{\da}_{\bS}$ for any density operator $\rho_{\bS}$, if and only if $F(\cE_{\al},U_{\bS})=1$. However, it has been shown that for several gates there is a constraint on the implementation naturally imposed by conservation laws \cite{02CQC, 03UPQ, 07CLI}. In the present paper, we assume that there are additively conserved quantities $L_{\bS}$ and $L_{\bA}$ of the systems $\bS$ and $\bA$, respectively, 
so that the unitary operator $U$ should satisfy the conservation law
\beqa\label{eq:conservation}
[U,L]=0,
\eeqa
where $L=L_{\bS}\otimes I_{\bA}+I_{\bS}\otimes L_{\bA}$, 
and $\ia$ is the identity operator of $\bA$.

Since different pairs of observables can represent the same 
additive conservation law,
it is convenient to introduce a standard representation of the conserved 
quantity.
Since a scalar operator poses no constraint, we always assume that 
$L_{\bS}$ has two distinct real eigenvalues $a<b$.
Let $c=(b-a)/2$.  Then, it is easy to see that $c$ is the maximum
standard deviation of $L_{\bS}$ attained, for instance, by 
$\ket{\ps}=2^{-1/2}(\ket{L_{\bS}=a}+\ket{L_{\bS}=b})$.
It is also easy to see that the operator 
$c^{-1}(L_{\bS}-bI_{\bS})+I_{\bS}$ is a non-scalar, 
unitary and self-adjoint operator.  Therefore this operator can be represented as
\beqa
c^{-1}(L_{\bS}-bI_{\bS})+I_{\bS} &=& \vec{l}\cdot\vec{\si} \label{eq}
\eeqa
with a real vector $\vec{l}=(l_x,l_y,l_z)$ satisfying $\| \vec{l}\|^2=l_x^2+l_y^2+l_z^2=1$. Equivalently we have
\beqa
L_{\bS} &=&(b-c)I_{\bS}+c\,\vec{l}\cdot\vec{\si}. 
\label{eq:7}
\eeqa
Here $b$ and $c$ are the maximum eigenvalue and the maximum standard 
deviation of $L_{\bS}$, respectively.
The addition of a scalar operator to $L_{\bS}$ does not affect the condition \Eq{conservation}, and it also does not change the standard deviation of $L_{\bS}$; 
thus, the pair $(L_{\bS},L_{\bA})$ and the pair $(c\,\vec{l}\cdot\vec{\si},L_{\bA})$
represent the same additive conservation law.
From the above, it is also true that the pair $(\vec{l}\cdot\vec{\si},L_{\bA}/c)$
represents the same conservation law as the pair $(L_{\bS},L_{\bA})$.
We shall use such simplifications where it is useful.

As above, the conserved quantity $\ls$ is determined by the vector $\vl$, which  can be parameterized as $\vec{l}=(\sin\omega \cos\chi, \sin\omega \sin\chi, \cos\omega)$ with $0 \leq \omega \leq \pi$ and  $0 \leq \chi < 2\pi$. As a reference, we introduce a unit vector $\vec{e}=(0,0,1)$, which is the vector $\vl$ for $\omega=\chi=0$. Then, there exists a unitary transformation $R_{\bS}(\vl)$ of the system $\bS$ such that 
\beqa
L_{\bS}(\vec{l}) &=& {R_{\bS}(\vl)}^{\da}\ls(\vec{e})R_{\bS}(\vl)
\eeqa
for any $\vec{l}$.
For instance, we can take ${R_{\bS}(\vl)}=\exp[-i\omega(\vec{n}\cdot\vec{\sigma})/2]$, where $\vec{n} = (\vec{l}\times \vec{e})/\| \vec{l}\times \vec{e} \|  = (\sin\chi, -\cos\chi,0)$ so that we have $R_{\bS}(\vl) =\cos(\omega/2)\is -i \sin(\omega/2)(\sin\chi\xs - \cos\chi\ys)$. Now, we assume that $L_{\bA}$ also depends on $\vl$ and that there is a unitary transformation $R_{\bA}(\vl)$ such that
\beqa
L_{\bA}(\vec{l}) &=& {R_{\bA}(\vl)}^{\da}L_{\bA}(\vec{e})R_{\bA}(\vl)
\label{unitary-rotation}
\eeqa
for any $\vec{l}$. 
Then, the conserved quantity of the system $\bS+\bA$ is transformed as
$$L(\vec{l}) = R(\vl)^{\da}L(\vec{e})R(\vl),$$ 
where $R(\vl)=R_{\bS}(\vl) \otimes R_{\bA}(\vl)$ and $L(\vl) = L_{\bS}(\vec{l}) \otimes \ia + \is \otimes L_{\bA}(\vec{l})$.

The above formulation typically includes the following examples.

{\em Angular momentum conservation law}.
In this case, we assume that the system $\bS$ has the spin $(\hbar/2)\vec{\si}$.
Then, $L_{\bS}=(\hbar/2)\vec{l}\cdot\vec{\si}$ represents 
the spin component along the
$(\omega,\chi)$ direction.  The angular momentum conservation law 
in the $(\omega,\chi)$
direction is represented as the case where $L_{\bA}$ is 
the angular momentum of $\bA$ in the $(\omega,\chi)$ direction.
In this case, we have $b=c=\hbar/2$ and $\vec{l}$ is arbitrary.
 
{\em Atom-field interaction}.
An additive conservation law holds for the well-known Jaynes-Cummings 
model 
\cite{JC63},
which describes the coupling of a two-level atom $\bS$ 
with a single-mode $\bA$ of the
electromagnetic field with annihilation operator $a$.  
Allowing for a detuning $\Delta$ between
the atom and the field, the Hamiltonian for the model may be written in a
suitable interaction picture as
\begin{equation}
H = \hbar\Delta I\otimes a^\dagger a 
+ i\hbar g \left(\ket 0 \bra 1\otimes a - \ket 1 \bra 0 \otimes a^\dagger \right),
\label{twelve}
\end{equation}
where $g$ is an appropriate coupling constant. 
Then, $L_{\bS}=Z_{\bS}$ and $L_{\bA}=2a^\dagger a$ 
constitute a pair of additively 
conserved quantities for $U=e^{-itH/\hbar}$ for any real $t$ \cite{05CQL}.
Thus, the constraint applies when one wants to realize a single-qubit  
gate by the Jaynes-Cummings interaction with the
parameters $\Delta$, $g$, and $t$. 
In this case, we have $b=c=1$ and $\vec{l}=(0,0,1)$.
For multimode fields, we refer the reader to Refs.~\cite{05CQL,06MEP}.

\section{Mean square deviation \label{Perror-v}}

In order to obtain an upper bound of the gate fidelity $F(\cE_{\al},U_{\bS})$,
we shall study the gate infidelity 
defined as
$1-F(\cE_{\al},U_{\bS})^2$,
and show that an additive conservation law generally poses a lower bound 
on the gate infidelity.
It is expected that such a limitation can be derived by the uncertainty 
relation generally formulated by Robertson \cite{Rob29} on 
quantum fluctuations, measured by standard deviations or variances,
of arbitrary pairs of non-commuting observables.
In fact, in Refs.~\cite{02CQC,03UPQ} commutation relations satisfied by 
noise and disturbance operators with the conserved quantity have been 
considered in order to apply Robertson's inequality, and lower bounds have been
obtained for the gate infidelities of  the CNOT gate and the Hadamard gate.
Here, we extend the above method to arbitrary single-qubit gates.

For this purpose, we introduce the deviation operator $D$ of the system
$\bS+\bA$ defined by 
\begin{eqnarray}\label{eq:DO}
D= U^{\da}(L_{\bS}\otimes I_{\bA})U 
- {U_{\bS}}^{\da} L_{\bS} U_{\bS}  \otimes I_{\bA}, 
\label{Deviation_theta-phi}
\end{eqnarray}
and we shall show that the variance of the deviation
operator is, up to a constant factor, a lower bound on the gate infidelity.
By the obvious cancellation when Eq.~(\ref{eq:7}) is substituted in Eq.~(\ref{eq:DO}), we  can assume $L_{\bS}=c\,\vec{l}\cdot\vec{\si}$ without
any loss of generality.
Then, $L_{\bS}$ has eigenvalues $\pm c$.
Let $\left| \chi_{0}  \right\rangle $ and $\left| \chi_{1}  \right\rangle $ be unit eigenvectors of $L_{\bS}$
with eigenvalues $c$ and $-c$, respectively.  
We define an orthonormal basis $\{ | \xi_{0}  \rangle $, $ | \xi_{1} \rangle \}$ as 
$|  \xi_{0}   \rangle = {U_{{\bS}}^{\dagger}}
\left|  \chi_{0} \right\rangle $ and $
| \xi_{1}  \rangle = U_{{\bS}}^{\dagger} \left| \chi_{1} 
\right\rangle$.
For the input states $\left| \xi_{i} \right\rangle$ with $i=0,1$,
the fidelity is given by
\begin{eqnarray}
F(| \xi_{i} \rangle )
&=& \left\langle \chi_{i} \right|
{\mathcal{E}}_{\al} \left(   \left| \xi_{i} \right\rangle
\left\langle \xi_{i} \right| \right)  
\left| \chi_{i} \right\rangle^{1/2}.
\label{basisFid.}
\end{eqnarray}
Since ${\mathrm{Tr}} [{\mathcal{E}}_{\al}( |\xi_{i} \rangle\langle 
\xi_{i} | ) ] =1$, we obtain
\begin{eqnarray}
\left\langle \chi_{i} \right|
{\mathcal{E}}_{\al} \left(   \left|  \xi_{j} \right\rangle
\left\langle  \xi_{j} \right| \right)  
\left| \chi_{i} \right\rangle
&=&1 - F(| \xi_{j} \rangle )^{2},
\label{1-fid.}
\end{eqnarray}
where $i \neq j$ for $i,j=0,1$.
The left-hand side can be described by using the general description of the output state $U (\left| \xi_{i} \right\rangle \otimes \left| A \right\rangle )$ of the system $\bS + \bA$ as
\begin{eqnarray}
U (| \xi_{i} \rangle \otimes | A \rangle )
&=& | \chi_{0} \rangle \otimes | A^{i}_{0} \rangle 
+ | \chi_{1} \rangle \otimes |  A^{i}_{1} \rangle, 
\label{unitary}
\end{eqnarray}
where $| A^{i}_{j} \rangle $ are unnormalized states of $\bA$.	It is clear that\beqa
\left\langle \chi_{i} \right|
{\mathcal{E}}_{\al}  \left(   \left| \xi_{j} \right\rangle
\left\langle \xi_{j} \right| \right)  
\left| \chi_{i} \right\rangle
= \| \, |  A^{j}_{i} \rangle  \|^{2}.
\eeqa
We, therefore, have
\begin{eqnarray}
\|\, \ket{A^{j}_{i}} 
\|^{2} 
= 1 - F \left(  \left| \xi_{j} \right\rangle \right)^{2}
\label{E^2=1-F^2}
\end{eqnarray}
for $i \neq j$.

This fidelity $F \left(  \left| \xi_{j} \right\rangle \right)$ is related to  the mean square $\bracket{D^2}$ as follows.
The mean square in the state $| \psi \rangle \otimes | A \rangle$ is written by
\begin{eqnarray}
\langle D^{2} \rangle 
& =& \|\, [
(L_{\bS}  \otimes I)U - U
( U_{{\bS}}^{\dagger}L_{\bS}  U_{{\bS}}
\otimes I)
]
(| \psi \rangle \otimes | A \rangle)
\|^{2}. 
\label{DB^2}
\end{eqnarray} 
Here, as any input state $| \psi \rangle$ of $\bS$ can be described as 
$| \psi \rangle  = \cos\zeta\, | \xi_{0} \rangle
+e^{i\delta} \sin\zeta\, | \xi_{1} \rangle$,
where $0 \leq \zeta 
\leq \frac{\pi}{2}$ and $0 \leq \delta < 2 \pi$, and 
$L_{\bS}   = c| \chi_{0} \rangle \langle \chi_{0} | - c| \chi_{1} 
\rangle \langle \chi_{1} |$, from Eq.~(\ref{unitary}) we have 
\begin{eqnarray}
\langle D^2 \rangle
&=&
4c^2 (\| |A^{0}_{1} \rangle \| ^{2}\cos^{2}\zeta
+  \| |A^{1}_{0} \rangle \| ^{2} \sin^{2}\zeta). 
\eeqa
We substitute Eq.~(\ref{E^2=1-F^2}) into this equation and obtain
\beqa
\langle D^2 \rangle&=&4c^2\left[ 1 -
  F(|\xi_{0}\rangle)^{2}\cos^{2}\zeta 
- F(|\xi_{1}\rangle)^{2} \sin^{2}\zeta  \right]
\nonumber\\
&\leq&
4c^2 [ 1 - F({\mathcal{E}}_{\al},U_{\bS})].
\label{D^2=4(1-F^2)}
\end{eqnarray}
Since $\si(D)^{2}\le\bracket{D^2}$, where $\si(D)^2$ stands for the
variance of $D$ in the state $\ket{\ps}\otimes\ket{A}$,
we obtain 
\begin{eqnarray}\label{eq:I>V}
1-F({\mathcal{E}}_{\al},U_{\bS})^2
\geq
\frac{1}{4c^2}\sup_{\ket{\ps}}\si(D)^{2},
\label{b_of_infid}
\end{eqnarray}
where $\ket{\ps}$ varies over all the state vectors of $\bS$.

\section{Uncertainty relation \label{review02}}

In what follows, we shall consider the uncertainty relation between 
the deviation operator $D$ and the conserved quantity $L$ to obtain
an lower bound of the variance of $D$.

By the conservation law \eq{conservation}, we have
\begin{eqnarray} \label{eq:[DL]}
[ D,L ]
&=& - [ {U_{\bS}}^{\dagger}
L_{\bS} U_{\bS} , L_{\bS}  ]   \otimes I_{\bA}.
\end{eqnarray}
Since addition of a scalar operator does not affect the commutator, we assume 
$b=c$ or $L_{\bS}=c\,\vec{l}\cdot\vec{\si}$ without any loss of generality.  
By Robertson's inequality \cite{Rob29}, we have
\begin{eqnarray}
\si( D )\si (L) 
\geq
\frac{1}{2}| \langle [ D, L]  \rangle |, 
\label{Robertson}
\end{eqnarray}
where $\bracket{\cdots}$ and $\si$ stands for the mean value and standard deviation in the
state $\ket{\ps}\otimes\ket{A}$, respectively.
Since the state $\ket{\ps}\otimes\ket{A}$ is a product state, the variance of the conserved quantity is given by 
\beqa
\si(L)^{2}=\si(L_{\bS})^{2}+\si(L_{\bA})^{2}.
\eeqa
Thus, we have
\begin{eqnarray}
\si( D)^{2}
\ge
\left(\frac{1}{4}\right)
\frac{|\bracket{[U_{\bS}^{\dagger} L_{\bS}U_{\bS}, L_{\bS}]}|^2} 
{\si(L_{\bS})^{2}+\si(L_{\bA})^{2}}.
\end{eqnarray}
As the variance of $\ls$ is upper bounded by the maximum eigenvalue $c$, namely, $\si(L_{\bS})^2 \le c$, we see that
\begin{eqnarray}
\si( D )^{2}
\geq 
\left(\frac{1}{4c^2}\right)
\frac{|\langle [ {U_{\bS}}^{\dagger} L_{\bS} U_{\bS}, L_{\bS}  ] 
\rangle |^{2} }{1+\si(L_{\bA}/c)^{2} } 
\label{VV>}
\end{eqnarray}
for any $\ket{\ps}$ and $\ket{A}$.
We are interested in the worst case error $\sup_{\ket{\ps}}\si(D)^{2}$.
Taking the supremum over $\ket{\ps}$ of the both sides, and noting
the relation
$\sup_{\ket{\ps}}|\bracket{X}|= \|X\|$ for any operator on $\cH_{\bS}$,
we obtain
\beqa
\sup_{\ket{\ps}}\si(D)^{2} 
\geq 
\left(\frac{1}{4c^2}\right)
\frac{\| \,[ {U_{\bS}}^{\dagger} L_{\bS} U_{\bS}, L_{\bS}  ]\, \|^{2} }
{1+\si(L_{\bA}/c)^{2}} 
\label{b_of_sup}
\eeqa
for any $\ket{A}$.

\section{Constraint with relative angle in the general case \label{Relative-angle}}

In what follows, we determine the operator norm
$
\| \,[ {U_{\bS}}^{\dagger} L_{\bS} U_{\bS}, L_{\bS}  ]\, \|
$
in geometrical terms from the vector analysis in three-dimensional space.

To calculate this operator norm, recall the relations $L_{\bS}/c=\vec{l}\cdot\vec{\si}$ and $U_{\bS}=e^{i\phi}(\cos\frac{\theta}{2}\is + i\sin\frac{\theta}{2} \vec{u}\cdot\vec{\si})$, and calculate
\beqa
( \vl \cdot \vs) (\vu\cdot \vs) =
(\vl \cdot \vu )\is + i (\vl\times \vu) \cdot \vs,
\label{l-by-u}
\eeqa
where $\vl \cdot \vu = l_{x}u_{x} + l_{y}u_{y} + l_{z}u_{z}$ and  
$ \vl \times \vu
 = (l_{y}u_{z}-l_{z}u_{y}, l_{z}u_{x}-l_{x}u_{z}, l_{x}u_{y}-l_{y}u_{x})$.
We are going to use Eq.~(\ref{l-by-u}) repeatedly. A straightforward calculation leads to
\beqa
\hspace{-5mm}
\us^{\dagger}\ls \us 
= \cos^2\frac{\theta}{2}\ls-2c\sin\frac{\theta}{2}\cos\frac{\theta}{2}(\vl\times\vu)\cdot\vs 
+\sin^2\frac{\theta}{2} (\vu\cdot\vs)\ls(\vu\cdot\vs).
\eeqa
Now take the commutator of this with $\ls$. The first term gives zero, and the last one may be written as
\beqa
\hspace{-18mm}
\left[\sin^{2}\frac{\theta}{2} (\vu\cdot \vs) L_{{\bS}}(\vu\cdot \vs), 
L_{{\bS}}\right] 
= 2i c^2 \sin^{2}\frac{\theta}{2} \left[( \vl \cdot \vu ) (\vu \times \vl) - 
\{ \vu \times (\vl \times \vu) \} \times {\vl} \right]\cdot \vs.
\eeqa
As $\vu$ and $\vl$ are unit vectors, the vector analysis gives 
$
 \big( \vu \times ( \vl \times \vu ) \big) \times \vl 
= - (\vl \cdot \vu ) (\vu  \times \vl ).  
$
Thus we have
\begin{eqnarray}
[ (\vu\cdot \vs) L_{{\bS}}(\vu\cdot \vs) , 
L_{{\bS}}] 
&=& 4c^2i   (\vl \cdot \vu) (\vu \times \vl)  \cdot \vs,
\end{eqnarray}
and accordingly,
\beqa
[\us^{\dagger}\ls\us, \ls]\cr
=4ic^2\sin\frac{\theta}{2}\left[ \cos\frac{\theta}{2}\left\{ \left(\vu \times \vl\right) \times \vl \right\} 
+ \sin\frac{\theta}{2} \left(\vu\cdot\vl\right)\left(\vu\times\vl\right) \right]  \cdot \vs.\label{ulu_l}
\eeqa
In Eq.~(\ref{ulu_l}), the vectors $(\vu \times \vl) \times \vl$ of the first term and $(\vu\times\vl)$ of the second term are mutually orthogonal and have the same norm $\| \vu \times \vl \|$. Hence the square of the operator norm of the commutator can be written by 
\beqa
\|[\us^{\dagger}\ls\us, \ls]\|^2 =
16c^4 \sin^2\frac{\theta}{2} \| \vu \times \vl \|^2 \left\{ \cos^2 \frac{\theta}{2}+ \sin^2\frac{\theta}{2} \left(\vu\cdot\vl \right)^2\right\},
\eeqa
because any vector $\vec{a}$ the operator norm  $\|\vec{a} \cdot \vs \|$ is given by $\| \vec{a} \|.$

Here, the vectors $\vu$ and $\vl$ have the following geometric meaning
with respect to the Bloch sphere determined by $\vec{\si}$.  We note that, in general, an operator of the form $\exp(-i\theta (\vec n\cdot\vec\sigma)/2)$ causes a rotation of the qubit state in the Bloch sphere by an angle $\theta$ around an axis given by the unit vector $\vec n$; by the representation (\ref{eq2}) and (\ref{eq4}), we see that the desired gate operation $U_{\bS}$ is, up to a phase factor, precisely such a rotation, by an angle $\theta$, around an axis given by $\vec u$. 
On the other hand,  the vector $\vl$ represents the direction of the conserved quantity (for instance, the direction of a conserved angular momentum component), and therefore is an axis of symmetry of the system, since a rotation around $\vl$, given by $R_L(t)=\exp(-i t L_{\bS}/2c)$, commutes with $U$. 
Let $\Psi$ be the relative angle between those two axes 
corresponding to $R_U$ and $R_L$.
Then, $\Psi$ is characterized by 
\beqa
\vl \cdot \vu=\cos\Psi\quad\mbox{and}\quad
\|\vu \times \vl \|=\sin\Psi.
\eeqa
Thus, we have
\beqa
\|\, [ U_{\bS}^{\dagger}L_{{\bS}}U_{\bS}, 
L_{{\bS}}] \,\|^2
=16c^4 \sin^2\frac{\theta}{2} \sin^2\Psi \left(1-\sin^2\frac{\theta}{2} \sin^2\Psi\right).
\label{op-norm}
\eeqa
We combine the above relation with Eqs.~(\ref{b_of_infid}) and (\ref{b_of_sup}), and obtain 
\begin{eqnarray}\label{eq:CQL}
1-F({\mathcal{E}}_{\al},U_{\bS})^2
\geq 
\frac{ \sin^2\frac{\theta}{2} \sin^2\Psi (1-\sin^2\frac{\theta}{2} \sin^2\Psi)}{1+\si(L_{\bA}/c)^{2}}. \label{general-bound1}
\end{eqnarray}
We have thus successfully generalized the result in Ref.~\cite{03UPQ} for the Hadamard gate to all the single-qubit gates. To see this, when $\theta= \pi$ (corresponding to a self-adjoint gate $U_{\bS}$), we have
\begin{eqnarray}
1-F({\mathcal{E}}_{\al},U_{\bS})^2
\geq 
\frac{ \sin^2 2\Psi }{4(1+\si(L_{\bA}/c)^{2})}.
\label{bound-for-self-adjoint} 
\end{eqnarray}
For $L_{\bS}=Z_{\bS}$ or $L_{\bS}=X_{\bS}$ the relative angle of the Hadamard gate is $\pi/4$, and the lower bound becomes ${4(1+\si(L_{\bA}/c)^{2})}^{-1}$, which was proved in the paper \cite{03UPQ}.  
When $\theta=0$, the operator $\us$ is proportional to the identity, and, of course, no restrictions apply in that case.

\begin{figure}
\centering
\includegraphics[width=8cm,clip]{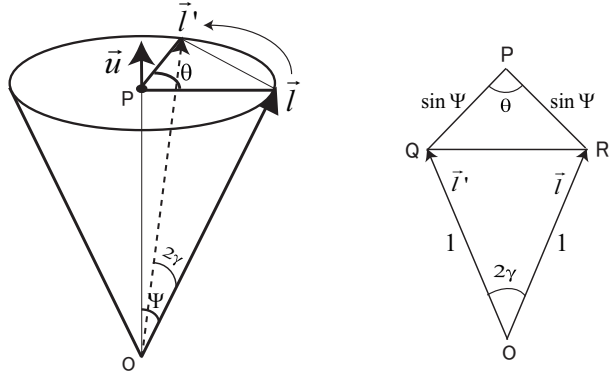}
\caption{The left figure describes the rotation of the vector $\vl$ in the Bloch sphere representation. The vector $\vl$ rotates about the axis vector $\vu$ by an angle $\theta$, which corresponds to the unitary transformation $\us^{\dagger}\ls\us$. The rotated vector $\vl'$ and the original vector $\vl$ make a relative angle $2 \gamma$ with each other.  The right figure is just the interior elevation.}
\label{fig1}
\end{figure}

Eq.~(\ref{op-norm}) has a simple geometrical representation. See the left figure of Fig.~\ref{fig1}. In the Bloch sphere representation, $\us$ is the rotation about the axis vector $\vu$ by an angle $\theta$. By the unitary transformation of the conserved quantity, i.e., $U_{\bS}^{\dagger}L_{{\bS}}U_{\bS}$, the vector $\vl$ is being rotated to the vector $\vl'$ defined by $U_{\bS}^{\dagger}L_{{\bS}}U_{\bS} = c \vl' \cdot \vs$.  Since $\| [\vec{a}\cdot \vs, \vec{b}\cdot \vs] \|$ = 2 $\| \vec{a} \times \vec{b} \|$ for any two self-adjoint operators $\vec{a}\cdot \vs$ and  $\vec{b}\cdot\vs$, we have 
$
\| [U_{\bS}^{\dagger}L_{{\bS}}U_{\bS}, \ls] \| = 2 c^2\| \vl' \times \vl \|.
$
Let the relative angle between the rotated vector $\vl'$ and the vector $\vl$ be $2 \gamma$, that is, $\| \vl' \times \vl \| = \sin 2\gamma$. We rewrite the square of the operator norm as
\beqa
\| [U_{\bS}^{\dagger}L_{{\bS}}U_{\bS}, \ls] \|^2 = 4c^4  \sin^2 2\gamma.
\label{op-norm2}
\eeqa
This equation must be consistent with Eq.~(\ref{op-norm}). This is easily seen from the right figure of Fig. \ref{fig1}. Suppose that the point Q and R are located  at the tips of the vector $\vl'$ and $\vl$, respectively. The length of segment QR is described by $\sin \gamma$, and at the same time it can be described by $2\sin\Psi \sin\frac{\theta}{2}$ because PQ=PR=$\sin\Psi$. Therefore we see that $ \sin^2 2\gamma = 4\sin^2 \frac{\theta}{2} \sin^2\Psi (1- \sin^2 \frac{\theta}{2}\sin^2\Psi)$, yielding Eq. (\ref{op-norm}) from Eq. (\ref{op-norm2}).

In terms of the relative angle $2 \gamma$, the lower bound of the gate infidelity is written by 
\begin{eqnarray}\label{eq:CQL}
1-F({\mathcal{E}}_{\al},U_{\bS})^2
\geq 
\frac{\sin^2 2\gamma }{4(1+\si(L_{\bA}/c)^{2})}. \label{general-bound1-1}
\end{eqnarray}
If $\theta=\pi$, it is easily seen from the left figure of Fig.~\ref{fig1} that $2\gamma$ becomes $2\Psi$. In this case, we obtain the lower bound (\ref{bound-for-self-adjoint}) for any self-adjoint gate. This is expected because any self-adjoint gate is the rotation about the axis $\vu$ by an angle $\pi$.

\section{Constraint under the angular momentum conservation law}

Now, we consider the case in which all components of the angular momentum are conserved, which means that $U$ must have full rotational invariance.
In this case, for the given $U_{\bS}$ the conservation law
\eq{conservation} holds for any direction $(\omega,\chi)$, so that
 (\ref{general-bound1}) holds for any $\Ps$ with $c=\hbar/2$. We will derive a lower bound of the gate infidelity maximized over all the relative angles, which gives the lower bound under rotational symmetry.  
To get a lower bound, which holds for any initial state $\ket{A}$, 
we note that for any given ancilla system $\bA$, 
the standard deviation $\si(2L_{\bA}/\hbar)$ is upper bounded by 
the operator norm $\|2L_{\bA}/\hbar\|$, if it is finite. 
We obtain
\begin{eqnarray}
1-F({\mathcal{E}}_{\al},U_{\bS})^2
\geq 
\frac{ \sin^2\frac{\theta}{2} \sin^2\Psi (1-\sin^2\frac{\theta}{2} \sin^2\Psi)}{1+ \| 2L_{\bA}(\vl)/\hbar \|^{2}} \label{eq:42}
\end{eqnarray}
for any initial state $\ket{A}$. Then the denominator of this lower bound is invariant over all the vectors $\vl$, because the operator norm is invariant under the unitary transformation defined by Eq.~(\ref{unitary-rotation}).

The numerator can be maximized by the proper choice of the vector $\vl$ specifying the relative angle $\Psi$. Since the quantity $\sin^2\frac{\theta}{2} \sin^2\Psi (1-\sin^2\frac{\theta}{2} \sin^2\Psi)$ is upper bounded by $(\sin^2 \theta)/4$ for $0 \leq \theta \leq \pi/2$, and by $1/4$ for $\pi/2 \leq \theta \leq \pi$, 
 we obtain
\beqa\label{eq:CQL2a}
1-F({\mathcal{E}}_{\al},U_{\bS})^2
\geq 
\frac{\sin^2 \theta}{4(1+\| 2L_{\bA}(\vl)/\hbar \|^{2})}
\eeqa
for $0 \leq \theta \leq \pi/2$, and 
\beqa\label{eq:CQL2b}
1-F({\mathcal{E}}_{\al},U_{\bS})^2
\geq 
\frac{1}{4(1+\| 2 L_{\bA}(\vl)/\hbar \|^{2})}
\eeqa
for $\pi/2 \leq \theta \leq \pi$.
The lower bound decreases as the operator norm of the conserved
quantity increases.  
Thus, if $\bA$ is a spin
$N/2$ system we have $\|2L_{\bA}/\hbar\|=N$ and hence
\beqa\label{eq:CQL2c}
1-F({\mathcal{E}}_{\al},U_{\bS})^2
\geq 
\frac{\sin^{2}\theta}{4(1+ N^2)}
\eeqa
for $0 \leq \theta \leq \pi/2$, and 
\beqa\label{eq:CQL2d}
1-F({\mathcal{E}}_{\al},U_{\bS})^2
\geq 
\frac{1}{4(1+ N^2)}
\eeqa
for $\pi/2 \leq \theta \leq \pi$.
In particular, if the ancilla consists of $N$ qubits, that is, the total spin number is $N/2$,  we  can conclude that any single-qubit gate on a spin 1/2 qubit system cannot be implemented within the gate infidelity less than $(4+4 N^2)^{-1}$ (up to the constant $\sin^{2}\theta$ for $0 \leq \theta \leq \pi/2$) 
by a rotationally invariant interaction with an $N$ qubit ancilla system, 
or with any ancilla with spin $N/2$.

\section{An alternative lower bound \label{alternative}}

The derivation in Section \ref{Relative-angle} (in particular,
Eq.~(\ref{eq:CQL})) suggests that there is no constraint on the
realization of gates for which
$\Psi=0$ (that is, $\vec{u}=\vec{l}$) or  
$\Psi=\pi/2$ (that is, $\vec{u} \cdot \vec{l} = 0$) and $\theta=\pi$.  While the first conclusion is correct, the second one is not true in general, since, for instance, a lower bound for the gate trace distance $D(\cE_{\al},U_{\bS})$ has been obtained recently \cite{07CLI} 
for the bit flip 
or quantum NOT gate, $U_{\mathbf{S}}=X_{\mathbf{S}}$, under $b=c=1$ and $\vec{l}=(0,0,1)$.
However, since the gate infidelity is always dominated by the gate trace distance,
i.e., $D(\cE_{\al},U_{\bS})\ge 1-F(\cE_{\al},U_{\bS})^{2}$, we cannot
immediately derive constraints for the gate fidelity from the result in \cite{07CLI}.
Here, we shall instead show directly that such constraints do exist.

To motivate the calculations that follow, consider again the case just mentioned, with $U_{\mathbf{S}}=X_{\mathbf{S}}$ (i.e., $\vec u = (1,0,0)$, $\theta=\pi$, and $\phi=-\pi/2$), and a conservation law given by Eq.~(\ref{eq:7}) with $b=c=1$ and $\vec{l}=(0,0,1)$ (i.e., $L_{\mathbf{S}} = Z_{\mathbf{S}}$). The difficulty is that in this case the ideal gate operator $U_{\mathbf{S}}$ transforms $L_{\mathbf{S}}$ into something that commutes with itself:
\beqa
U_{\mathbf{S}}^\dagger L_{\mathbf{S}} U_{\mathbf{S}} = X_{\mathbf{S}} Z_{\mathbf{S}} X_{\mathbf{S}} = - Z_{\mathbf{S}} = - L_{\mathbf{S}}.
\eeqa
As a result, no constraint on $D$ follows from Eq.~(\ref{eq:[DL]}), since the commutator on the right-hand side vanishes.  However, the situation is different when we look at the action of $U_{\mathbf{S}}$ on other system operators, in particular $X_{\mathbf{S}}$ and $Y_{\mathbf{S}}$.  We have 
\begin{eqnarray}
U_{\mathbf{S}}^\dagger X_{\mathbf{S}} U_{\mathbf{S}} &=& X_{\mathbf{S}}, \nonumber \\
U_{\mathbf{S}}^\dagger Y_{\mathbf{S}} U_{\mathbf{S}} &=& -Y_{\mathbf{S}}, 
\end{eqnarray}
so if we define, as in Eq.~(\ref{Deviation_theta-phi}), the corresponding deviation operators
\begin{eqnarray}
D_x &=& U^{\da}(X_{\bS}\otimes I_{\bA})U - X_{\bS}  \otimes I_{\bA}, \nonumber \\
D_y &=& U^{\da}(Y_{\bS}\otimes I_{\bA})U + Y_{\bS}  \otimes I_{\bA},
\label{dxdy}
\end{eqnarray}
we find (using the conservation law, $[U,L]=[U^{\da},L]=0$)
\begin{eqnarray}
[D_y, L] &=& U^{\da}[Y_{\bS}\otimes I_{\bA},L]U + [Y_{\bS},L_{\bS}] \otimes I_{\bA} \nonumber \\
&=& 2i U^{\da}(X_{\bS}\otimes I_{\bA}) U  + 2i X_{\bS} \otimes I_{\bA}\nonumber \\
&=& 2i D_x + 4i X_{\bS}\otimes I_{\bA},
\label{eq:39}
\end{eqnarray}
which shows that $D_y$ and $D_x$ cannot simultaneously be zero.  Now, it is important to realize that the derivation of Eq.~(\ref{eq:I>V}) does not depend on $D$ being the deviation operator for the system's conserved quantity:  an identical lower bound on the fidelity could be obtained from the deviation operator of any system Hermitian operator with eigenvalues $\pm c$.  Accordingly, in what follows we show how a generalization of the result (\ref{eq:39}), to arbitrary $U_{\bS}$ and $L_{\bS}$, can be used in this way to derive a general lower bound for the gate infidelity that does not vanish when $\Psi = \pi/2$ and $\theta=\pi$, and thus in a sense complements the one obtained in Section 5, Eq.~(\ref{eq:CQL}).

To simplify the calculations that follow, we introduce three operators $\li$ defined as 
\beqa
\lth &=& \vl\cdot\vs, \cr
\ltw &=& \frac{1}{\sin\Psi}(\vl \times \vu)\cdot \vs, \cr
\lon &=& -i \ltw \lth =\frac{1}{\sin\Psi}\{(\vl \times \vu) \times \vl\} \cdot \vs = \frac{1}{\sin\Psi} \vu\cdot\vs - \frac{\cos\Psi}{\sin\Psi} \vl\cdot\vs. 
\label{rots}
\eeqa
These operators obeys the same commutation relation as the Pauli operators:
\beqa
[\li, \lj]=2i\sum_{k}\epsilon_{ijk}\lk,
\label{levi-civita}
\eeqa
where $\epsilon_{ijk}$ is the Levi-Civita symbol; in fact, they are essentially the Pauli operators, in a rotated reference frame. We define a new vector $\vs'=(\lon, \ltw, \lth)$. Because any self-adjoint, unitary operator can be written in terms of  $\vs'$ as  $\vec{a}\cdot\vs'$, where  $\vec{a}$ is a unit vector, we can write 
\beqa
\vu \cdot \vs = \vu' \cdot \vs'.
\eeqa
with a unit vector $\vu'= (u_{1},u_{2},u_{3})$. From the definition (\ref{rots}) it is clear that 
\beqa
u_{2}= \frac{1}{2}\Tr[\ltw (\vu' \cdot \vs')] = \frac{1}{2\sin{\Psi}}\Tr[\{(\vl\times\vu)\cdot \vs\}(\vu \cdot \vs)] =0.
\eeqa
Therefore, $\us$ may be written as 
\beqa
\us =e^{i\phi} \left\{ \cos\frac{\theta}{2} \is + i\sin\frac{\theta}{2} (\vu'\cdot \vs') \right\}
\eeqa
with $u_{2} = 0.$
Straightforward calculations then yield
\beqa
\hspace{-10mm}
\us^{\dagger} \li \us
= \cos^2\frac{\theta}{2}\li-2\sin\frac{\theta}{2}\cos\frac{\theta}{2}(\vl_{i}\times\vu')\cdot\vs' 
+\sin^2\frac{\theta}{2} (\vu'\cdot\vs')\li(\vu'\cdot\vs')
\eeqa
for $i=1,2$, where we define $\vl_{1}=(1,0,0)$ and $\vl_{2}=(0,1,0)$. Here in the third term,  $(\vu'\cdot\vs')\li(\vu'\cdot\vs')$ can be written as
\beqa
(\vu'\cdot\vs')\lon(\vu'\cdot\vs') 
= ({u_{1}}^2-{u_{3}}^2) \lon + 2u_{1} u_{3}  \lth
\eeqa
for $i=1$, and 
\beqa
(\vu'\cdot\vs')\ltw(\vu'\cdot\vs') 
= -({u_{1}}^2+{u_{3}}^2) \ltw
\eeqa
for $i=2$.
We calculate the commutator $[\us^{\dagger} \li \us, \lth]$
by using the commutation relation (\ref{levi-civita}) and obtain
\beqa
&&\hspace{-20mm}[\us^{\dagger} \lon \us, \lth] =2i \left\{ -\cos^{2}\frac{\theta}{2} \ltw
+ 2u_{3}\sin\frac{\theta}{2}\cos\frac{\theta}{2} \lon 
- \left({u_{1}}^2 -{u_{3}}^{2}\right)\sin^2\frac{\theta}{2} \ltw 
\right\}, 
\label{Ul1U}\\
&&\hspace{-20mm}[\us^{\dagger} \ltw \us, \lth] =2i \left\{ \cos^{2}\frac{\theta}{2}\lon + 2 u_{3}\sin\frac{\theta}{2}\cos\frac{\theta}{2} \ltw - \left( {u_{1}}^2 + {u_{3}}^2\right)\sin^2\frac{\theta}{2} \lon  	
\right\}. \nonumber\\
\label{Ul2U}
\eeqa

Let then $L_{\bS} = (b-c)I_{\bS} + c \vl \cdot \vs$ and define accordingly, by analogy with (\ref{dxdy}), the deviation operators
\begin{eqnarray}
D_i &=& U^{\da}(\li \otimes I_{\bA})U - U_{\bS}^{\da}\li U_{\bS}  \otimes I_{\bA}
\label{dxdy2}
\end{eqnarray}
for $i=1,2$.
Making use of the conservation law and the commutation relation (\ref{levi-civita}) as in Eqs.~(\ref{eq:39}) above, we obtain the relation
\beqa
\frac 1 c\big [D_1, L \big] 
&=& -2i U^{\dagger} (\ltw \otimes \ia)U 
- \left[ \us^{\dagger} \lon \us, \lth\right] \otimes \ia.
\label{d_1l}
\eeqa
As $L_{\bS}$ can be written in terms of $\vs'$ as 
$L_{\bS} = (b-c)I_{\bS} + c \vl' \cdot \vs'$ with $\vl'=(0,0,1)$, 
we have $u_{1}= \| \vl' \times \vu'\|= \sin\Psi$ and $u_{3}= ( \vl' \cdot \vu') = \cos\Psi$.  These yield  the following description of Eq.~(\ref{d_1l}) by
substituting Eq.~(\ref{Ul1U}), 
\beqa
\frac 1 c\big [D_1, L \big]  
= -2i \left( D_{2} + 2 \us^{\dagger}\ltw\us \otimes \ia + 2 \vec{v}\cdot \vs' \otimes \ia \right),
\eeqa
where $\vec{v}=(v_{1}, v_{2},v_{3})$ with 
\beqa
v_{1} &=& 2 \cos\Psi \sin\frac{\theta}{2} \cos\frac{\theta}{2}, \cr
v_{2} &=& \cos^{2}\Psi \sin^{2}\frac{\theta}{2} - \cos^2\frac{\theta}{2}, \cr
v_{3} &=& -\sin\Psi \sin\frac{\theta}{2} \cos\frac{\theta}{2}.
\eeqa
In a similar way, we have
\beqa
\frac 1 c\big [D_2, L \big]  
= 2i \left( D_{1} + 2 \us^{\dagger}\lon\us \otimes \ia + 2 \vec{w}\cdot \vs' \otimes \ia \right),
\eeqa
where $\vec{w}=(w_{1}, w_{2},w_{3})$ with 
\beqa
w_{1} &=& \cos^2\Psi \sin^2\frac{\theta}{2} - \cos^2\frac{\theta}{2}, \cr
w_{2} &=& -2 \cos\Psi \sin\frac{\theta}{2} \cos\frac{\theta}{2}, \cr
w_{3} &=& -\sin\Psi \cos\Psi \sin^2\frac{\theta}{2}.
\eeqa
Then, as in Section \ref{review02}, Robertson's uncertainty relation implies that: 
\beqa
\sigma(D_1)^{2} 
&\geq& 
\frac{
|\langle D_2 + 2 \us^{\dagger} \ltw \us \otimes \ia + 2\vec{v}\cdot\vs' \otimes \ia \rangle|^{2}
}{1+\sigma(L_{\bA}/c)^{2}}, \label{sigma_x} 
\eeqa
Now suppose we evaluate the numerator of the right-hand side in Eq.~(\ref{sigma_x}) in the eigenstate of $U_{\mathbf{S}}^\dagger \ltw U_{\mathbf{S}}$ corresponding to the eigenvalue $+1$.  
Let $\av{\cdot}^\prime$ denote the expectation values in this particular state.  We then have
\beqa
\langle (D_1)^{2} \rangle'
&\geq&
\frac{|2 +  \langle D_2 \rangle' +2
 \langle \vv \cdot \vs' \rangle' |^{2}}{1+\sigma(L_{\bA}/c)^{2}}.
\label{equ} 
\eeqa
Taking the square root of Eq.~(\ref{equ}), we have 
\beqa
\sqrt{ \langle (D_1)^{2} \rangle' }
&\geq&
\frac{2 - | \langle D_2 \rangle' |
 -2 | \langle \vv \cdot \vs' \rangle' |   }{\sqrt{1+\sigma(L_{\bA}/c)^{2}}}.
\eeqa
As $\sqrt{ \sup \langle X^{2} \rangle } \geq 
\sqrt{ \langle X^{2} \rangle }  \geq  |\langle X \rangle | $ holds for any observable $X$, we have
\beqa
\sqrt{\sup\langle (D_1)^{2}}\rangle &\geq&  
\frac{2-\sqrt{\sup\langle (D_2)^{2} \rangle} -2\| \vv \| }{\sqrt{1+\sigma(L_{\bA}/c)^{2}}},
\label{y}
\eeqa
In a similar way, we have a lower bound of $\sqrt{\sup\langle (D_2)^{2}}\rangle $ as 
\beqa
\sqrt{\sup\langle (D_2)^{2}}\rangle &\geq& 
\frac{2-\sqrt{\sup\langle (D_1)^{2} \rangle} -2\| \vw \| }{\sqrt{1+\sigma(L_{\bA}/c)^{2}}}. 
\label{x}
\eeqa
Adding Eqs. (\ref{y}) and (\ref{x}) gives
\beqa
\sqrt{\sup\langle (D_1)^{2} \rangle } + \sqrt{ \sup\langle (D_2)^{2} \rangle}
\geq \frac{1}{1+\sqrt{1+\sigma(L_{\bA}/c)^{2}}}\{ 4 - 2(\| \vv \| + \| \vw \| ) \}. \nol
\eeqa

In the same way as in Section \ref{Perror-v} (cf. Eq.~(\ref{D^2=4(1-F^2)})), it can be shown that the gate infidelity is bounded from below by the mean square of $D_i$, i.e., 
\begin{equation}
1-F({\cal E}_\alpha,U_{\mathbf{S}})^2 \ge \frac{1}{4c^2}\sup_{\ket\psi} \langle {(D_{i})}^2 \rangle.
\label{e2}
\end{equation}
Therefore, we obtain
\beqa
1-F(\cE_{\alpha}, U_{\bS})^{2} 
\geq 
\frac{1}{\left(1+\sqrt{1+\sigma(L_{\bA}/c)^{2}} \right)^{2}}
\left[1 - \frac{1}{2}(\| \vv \| + \| \vw \|) \right]^{2}. 
\label{general-bound2-1}
\eeqa
Here $\| \vv \|$ and $\| \vw \|$ can be written as
\beqa
\| \vv \| 
&=& \sqrt{\cos^2\frac{\theta}{2} + \cos^2\Psi \sin^2\frac{\theta}{2} \left(\cos^2\frac{\theta}{2} + \cos^{2}\Psi \sin^{2}\frac{\theta}{2}\right)},
\label{vec-v} \\
\| \vw \|
&=&\sqrt{\cos^2\Psi \sin^2\frac{\theta}{2} + \cos^2\frac{\theta}{2} \left(\cos^2\frac{\theta}{2} + \cos^{2}\Psi \sin^{2}\frac{\theta}{2}\right)}.
\label{vec-w}
\eeqa
This bound given by Eqs.~(\ref{general-bound2-1})-(\ref{vec-w}) is fairly complicated. A slight simplification, resulting in a somewhat less tight bound, can be obtained by using 
\begin{equation}
|\vv|+|\vw|\leq 2 \sqrt{ \frac{|\vv| + |\vw|}{2}}.
\end{equation}
This yields
\beqa
\hspace{-20mm}
1-F(\cE_{\alpha}, U_{\bS})^{2} \cr
\hspace{-20mm}\geq 
\frac{1}{\left(1+\sqrt{1+\sigma(L_{\bA}/c)^{2}} \right)^{2}} 
\left[ 1 - \frac{1}{\sqrt{2}}\sqrt{\left(1-\sin^2{\frac{\theta}{2}}\sin^2\Psi\right)\left(2-\sin^2\frac{\theta}{2}\sin^2\Psi\right)} \right]^{2}. 
\label{general-bound2-2}
\eeqa
which can also be written as 
\begin{equation}
1-F(\cE_{\alpha}, U_{\bS})^{2} \ge \frac{1}{\left(1+\sqrt{1+\sigma(L_{\bA}/c)^{2}} \right)^{2}} 
\left[1-\frac{1}{\sqrt{2}} |\cos\gamma|\sqrt{1+\cos^2\gamma} \right]^2
\end{equation}
in terms of the angle $\gamma$ introduced in Figure 1.

It does not seem possible to express the tighter bound given by Eqs.~(\ref{general-bound2-1})-(\ref{vec-w}) in terms of a single angle.  However, in the case $\theta= \pi$,  Eqs.~(\ref{general-bound2-1})-(\ref{vec-w}) do yield a very simple bound:
\beqa
\hspace{-5mm}
1-F(\cE_{\alpha}, U_{\bS})^{2} \geq 
\frac{1}{\left(1+\sqrt{1+\sigma(L_{\bA}/c)^{2}} \right)^{2}} 
\left[ 1- \frac{1}{2} |\cos\Psi|(1+ |\cos\Psi | ) \right]^2.
\eeqa
The right-hand side of this equation is the lower bound for  
an arbitrary self-adjoint gate, i.e., $\us^{\dagger}=\us$ (up to a phase factor); the Hadamard gate and the Pauli gates have this property.

\begin{figure}
\centering
\includegraphics[width=12cm,clip]{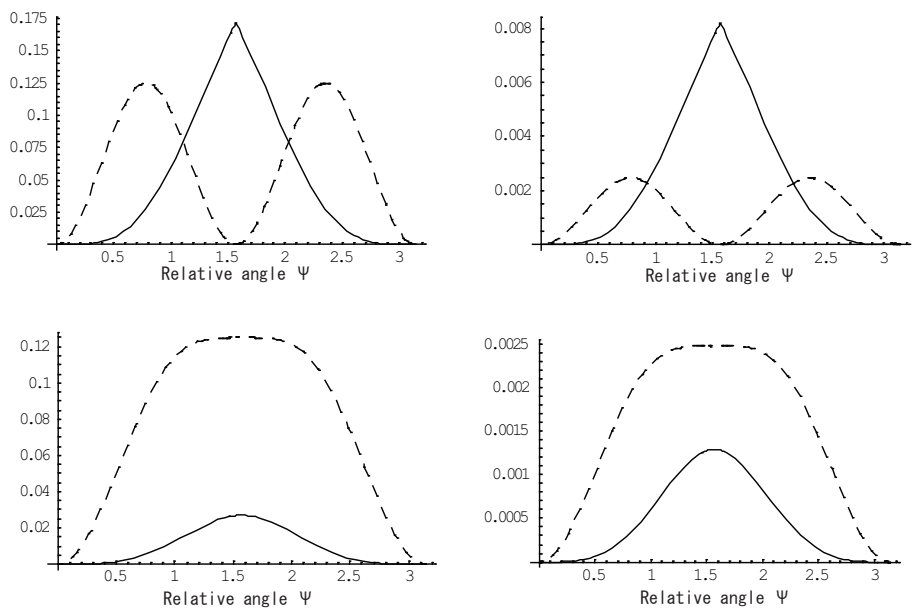}
\caption{Comparison between the lower bound in Eq.~(\ref{general-bound2-1}) (Solid line) and that in Eq.~(\ref{general-bound1}) (dashed line).  We set $\theta=\pi$ and $\sigma(L_{\bA}/c)=1$ in the upper left figure, and $\theta = \pi$ and $\sigma(L_{\bA}/c)=10$ in the upper right figure, and $\theta=\pi/2$ and $\sigma(L_{\bA}/c)=1$ in the lower left figure, and $\theta=\pi/2$ and $\sigma(L_{\bA}/c)=10$ in the lower right figure.}
\label{fig2}
\end{figure}

The general bound (\ref{general-bound2-1}) should be compared to the bound previously obtained by the first approach, that is, Eq.~(\ref{general-bound1}).
We plot these bounds both for $\theta= \pi$ and $\theta=\pi/2$ (See Fig.\ref{fig2}). Generally speaking, neither bound is tighter than the other. For $\theta=\pi/2$ and $\sigma(L_{\bA}/c)=1, {\rm{or}}\ 10$, the previous bound (\ref{general-bound1}) is tighter than the present bound (\ref{general-bound2-1}) over all the relative angle $\Psi$. For $\theta=\pi$, the present bound does not vanish when $\Psi = \pi/2$; in fact, it reaches its peak, whereas the previous bound vanishes, which was the main motivation for this section.

\section{Concluding remarks}
We investigated the limitations of the gate infidelity of implementing arbitrary single-qubit gates under arbitrary additive conservation laws. We obtained the two different lower bounds of the gate infidelity by using Robertson's uncertainty relation. Both bounds were described by the variance of the conserved quantity and the parameters $(\theta, \Psi)$. The parameter $\Psi$ is considered to be the relative angle between the axes of rotations on the Bloch sphere generated by the conserved  quantity and the gate operation, which provides a geometrical understanding of the limitations in the three dimensional space. For the first bound (\ref{general-bound1}), we show that it depends on the relative angle $2\gamma$ between the axis of rotations generated by the conserved quantity before and after the gate operation, in the Heisenberg picture. Both bounds are important because in general, neither bound is tighter than the other. Both bounds become zero if and only if $\theta=0$ or $\Psi=0$. This is expected because the conserved quantity $L$ commutes with the desired gate $\us$ only if $\theta=0$ or $\Psi=0$. Generally speaking, the limitations decrease as the variance of the conserved quantity increases, and given the variance and the parameters $\theta$ and $\Psi$, the upper bound of both bounds provides the fundamental limitation for any implementation of the desired gate under the conservation law.

\ack
This research was partially supported by the SCOPE project of the MIC, the  Grant-in-Aid
for Scientific Research (B)17340021 of the JSPS, and the CREST project of the JST.  J. G.-B.'s research was supported by the National Science Foundation.

\appendix 
\section{Ancilla in a mixed state \label{app}}
In this section, for completeness, we show that our results are not changed if we assume an arbitrary state $\rho_{\bA}$, which does not have to be a pure state, of the ancilla system as the initial state of the implementation $\cE_{\alpha}$.  In order to apply our previous treatment, we purify this state by introducing an auxiliary system $\bB$.  The dimension of the auxiliary system $\bB$ must be greater than or equal to that of the ancilla system so that the system $\bB$ can provide the purification of any state of the ancilla system $\bA$.  
Suppose that $\ket{A'}$ is the purified state for the state $\rho_{\bA}$ satisfying
\beqa
\Tr_{\bB}[\ketbra{A'}] = \rho_{\bA},
\eeqa
where $\Tr_{\bB}$ stands for the partial trace over the auxiliary system $\bB$.
$\ket{A'}$ is a vector of the Hilbert space of the composite system $\bA + \bB$. This state can be considered as the initial state of the ancilla system if the ancilla system is extended to the original ancilla system $ \bA$ plus the auxiliary system $\bB$. In this case, the total system that we have is the composite system $\bS + \bA + \bB$, and the initial state of the implementation is given by $\rho_{\bS}  \otimes \ketbra{A'}$. As the physical system is extended from the original system $\bS+\bA$ to the system $\bS + \bA + \bB$ to include the auxiliary system $\bB$, the time evolution operator of the implementation must also be extended to $U'= U \otimes I_{\bB}$, where $\ib$ is the identity operator of $\bB$. Therefore the implementation is given by the trace-preserving quantum operation defined as 
\beqa
\cE_{\alpha'} (\rho_{\bS}) = \Tr_{\bA+\bB}\left[ U'(\rho_{\bS}  \otimes \ketbra{A'})U'^{\dagger}  \right]
\eeqa
for any density operator $\rho_{\bS}$ of the system $\bS$, where $\alpha'$ stands for the characterization of this quantum operation, $\alpha' = (U', \ket{A'})$, and $\Tr_{\bA+\bB}$ stands for the partial trace over the Hilbert space of $\bA + \bB$. The evolution operator $U'$ must obey the conservation law
\beqa
[U', L'] =0, 
\eeqa
where $L'= \ls \otimes \ia \otimes \ib + \is \otimes L_{\bA} \otimes \ib$, which is the natural extension of the conserved quantity according to the above extension of the physical system. This conservation law is equivalent to the previous one $(\ref{eq:conservation})$.

It is not difficult to verify that this extension does not change the calculations leading to the general bounds, that is, we can apply the same arguments as from Sec.~3 to Sec.~7 even in this case, and obtain 
\begin{eqnarray}
1-F({\mathcal{E}}_{\alpha'},U_{\bS})^2
\geq 
\frac{ \sin^2\frac{\theta}{2} \sin^2\Psi (1-\sin^2\frac{\theta}{2} \sin^2\Psi)}{1+\si(L_{\bA}/c)^{2}}, \label{general-bound1-1}
\end{eqnarray}
and 
\beqa
1-F(\cE_{\alpha'}, U_{\bS})^{2} 
\geq 
\frac{1}{\left(1+\sqrt{1+\sigma(L_{\bA}/c)^{2}} \right)^{2}}
\left[1 - \frac{1}{2}(\| \vv \| + \| \vw \|) \right]^{2}. 
\label{general-bound2-1-1}
\eeqa
The only difficulty here is that the implementation $\cE_{\alpha'}$ appears different from the original one $\cE_{\alpha}$. However, this problem is removed because the implementation $\cE_{\alpha'}$ is equivalent to the quantum operation  characterized by the time evolution operator $U$ and arbitrary mixed states $\rho_{\bA}$ of the ancilla system, that is,
\beqa
\cE_{\alpha'} (\rho_{\bS}) = \Tr_{\bA}\left[ U (\rho_{\bS} \otimes \rho_{\bA}) U  \right]
\label{equiv_of_maps}
\eeqa
for any density operator $\rho_{\bS}$. The proof is as follows. Since $U'=U \otimes \ib$, we have 
\beqa
\cE_{\alpha'} (\rho_{\bS}) = \Tr_{\bA+\bB} \left[ U \otimes \ib (\rho_{\bS} \otimes \ketbra{A'}) U^{\dagger} \otimes \ib  \right].
\eeqa
As $\ketbra{A'}$ is the purification of the $\rho_{\bA}$, we see that 
\beqa
\cE_{\alpha'} (\rho_{\bS}) 
&=& \Tr_{\bA} \left[  U \Tr_{\bB} \left[ \rho_{\bS} \otimes \ketbra{A'} \right]  U^{\dagger}  \right] \cr
&=& 
\Tr_{\bA} \left[  U (\rho_{\bS} \otimes \Tr_{\bB} \left[ \ketbra{A'} \right])  U^{\dagger}  \right] \cr
&=& \Tr_{\bA}\left[ U (\rho_{\bS} \otimes \rho_{\bA}) U  \right],
\eeqa
which completes the proof. This proof has been already shown in Ref.~\cite{07CLI}. 

We can then substitute Eq.~(\ref{equiv_of_maps}) into Eq.~(\ref{general-bound1-1}) and~(\ref{general-bound2-1-1}) and obtain the general lower bounds, which are correct even if arbitrary initial states $\rho_{\bA}$ of the ancilla system are assumed: 
\begin{eqnarray}
1-F({\mathcal{E}}_{\alpha},U_{\bS})^2
\geq 
\frac{ \sin^2\frac{\theta}{2} \sin^2\Psi (1-\sin^2\frac{\theta}{2} \sin^2\Psi)}{1+\si(L_{\bA}/c)^{2}}, 
\end{eqnarray}
and 
\beqa
1-F(\cE_{\alpha}, U_{\bS})^{2} 
\geq 
\frac{1}{\left(1+\sqrt{1+\sigma(L_{\bA}/c)^{2}} \right)^{2}}
\left[1 - \frac{1}{2}(\| \vv \| + \| \vw \|) \right]^{2},
\eeqa
where the quantum operation $\cE_{\alpha}$ is defined as 
\beqa
\cE_{\alpha}(\rho_{\bS})= \Tr_{\bA}\left[ U (\rho_{\bS} \otimes \rho_{\bA}) U  \right]
\eeqa
for any state $\rho_{\bS}$.

\section*{References}


\end{document}